\documentstyle[epsf]{fbssuppl}

\title{Heavy-to-light form factors in the  quark  model with
heavy infrapropagators.}

\author{ M.A. Ivanov$^1$, T. Mizutani$^2$, and Yu.M. Valit$^1$}
\institute{$^1$Bogoliubov Laboratory of Theoretical Physics\\
Joint Institute for Nuclear Research\\
141980 Dubna (Moscow region), Russia \\
$\,$\\
$^2$Department of Physics\\
Virginia Polytechnic Institute and State University\\
Blacksburg, VA 24061} 
\begin{document}

\maketitle
\begin{abstract}
We calculate the heavy-to-light form factors in the relativistic quark
model with heavy infrapropagators. Their $q^2$-dependence in the
physical region is defined by two parameters: the "infrared" parameter $\nu$
characterizing the infrared behavior of the heavy quark and the mass
difference of the heavy meson and heavy quark $E=m_H-M_Q$.

It is found that the values of the $D\to K(K^*)$ and $D\to\pi(\rho)$
form factors at $q^2=0$ are in excellent agreement with the available
experimental data and other approaches whereas these values for
$B\to\pi(\rho)$ transitions are found to be larger than those of
several other models.

The obtained form factors are used to calculate the widths of
the semileptonic decays of B and D mesons. The comparison of
our results with the available experimental data and other
approaches is given.
\end{abstract}

\section{Introduction}

A theoretical study of the heavy-to-light transitions is more complicated than
heavy-to-heavy ones because of the heavy quark symmetry cannot be applied
in this case.
The results of nonperturbative calculations made within
lattice QCD \cite{BKS}-\cite{GSS},
QCD sum rules \cite{DP}-\cite{BBD}, and various quark models
\cite{ISGW}-\cite{BD} are seemed to be rather different each other
to give a reliable  conclusion about the behavior of the form factors
in the physical region. Nevertheless, such calculations may be useful
for our understanding what is really important under description
of the heavy-to-light transitions.

One has to remark that the most of mentioned approaches allows one to
calculate the values of the form factors at some fixed point of $q^2$,
the momentum transfer squared, and then to adopt some simple shape
for getting $q^2$ dependence in the full physical region.

The lattice calculations give the form factors near $q^2=q^2_{\rm max}$
and then extrapolate them to $q^2=0$ using polelike dependence.
The QCD sum rules method allows one to give the reliable results
at $q^2=0$ and then also the pole dominance hypothesis is used
for extrapolating them to physical region.
The $q^2$-dependence have been obtained
by using the QCD sum rules method in the paper \cite{BBD}. It was found
that the pole dominance dependence is a good approximation for the vector
form factors whereas this approximation is not applicable for the axial
form factors. In the nonrelativistic constituent quark model \cite{ISGW}
the form factors, computed with the overlap integral of the
nonrelativistic meson wave functions, have the exponential $q^2$ dependence.
The quasipotential approach \cite{FGM} predicts similar behavior.
The $q^2$ dependence of the relevant form factors was derived from the
independent quark model \cite{BD} with a scalar-vector-harmonic potential.
Some nonrelativistic \cite{AW} and relativistic \cite{BSW,KS,Wol} quark models
adopt polelike (monopole, dipole,...) ansatz for the $q^2$ dependence of
the form factors.

The dispersion relations incorporating with the Heavy Quark Effective Theory
(HQET) and  Chiral Perturbation Theory (ChPT) \cite{BK},
with the light-cone model  \cite{Mel},
with the crossing symmetry and QCD perturbative calculation
in unphysical kinematic region \cite{Bec} have been used
for studying the form factors.
In the paper \cite{Cas} heavy-light form factors have been analyzed
within approach incorporating HQET, ChPT and Vector Meson Dominance (VMD)
model.

Therefore it may be seen that a completely consistent calculation
of the heavy-to-light form factors in the full physical region
of the momentum transfer squared is a problem not only for the approaches
based on the first principles of QCD but also for various quark models.
The polelike $q^2$ dependence usually used for the form factors should
be at least justified from the more deep physical representations.

The present work is devoted to the calculation of the form factors describing
the heavy-to-light transitions within a relativistic quark model
with confined light quarks and infrared heavy quark propagators \cite{INFRA}.

Our previous approach \cite{IKM,EI} was based upon the quark confinement
model (QCM) which incorporates the confinement of light quarks by devising
a quark propagator that has no singularities, thus forbidding the production
of a free quark \cite{EI}. This model then allows one to perform covariant
calculations of Feynman diagrams with dressed light quark propagators free of simple
pole. It was shown that  the QCM approach successful in many static and
non-static properties of the light hadrons \cite{EI}.  In the paper
\cite{INFRA} the infrared behavior of the heavy quark has been taken
into account by modifying its conventional propagator in terms of a single
parameter $\nu$. The weak decay constants and the Isgur-Wise function
have been calculated.

Here we slightly modify the inclusion of the infrared behavior of
the heavy quark in a such way to provide  the conservation of local
properties of Feynman diagrams like gauge invariance and the Ward-Takachashi
identity. We investigate the dependence of the heavy-to-light form factors
on two parameters: the infrared parameter $\nu$ and the mass difference
of the heavy meson and heavy quark $E=m_H-M_Q$.

We  found that the values of the $D\to K(K^*)$ and $D\to\pi(\rho)$
form factors at $q^2=0$ are in excellent agreement with the available
experimental data and other approaches whereas these values for
$B\to\pi(\rho)$ transitions are found to be larger those of
several other models.

The obtained form factors are used to calculate the widths of
the semileptonic decays of B and D mesons.
We compare  our results with the available
experimental data and other approaches.

\section{Model}

Our approach is based on the effective interaction Lagrangian
which describes the transition of hadron into quarks.
For example, the transition of heavy meson $H$ into heavy $Q$
and light $q$ quarks may be described by
\begin{equation}
\label{la}
{\cal L}_{\rm int}\!(x)\!\!=\!\!g_H\! H\!(x)\!\!\! \int\!\!\! dx_1\!\!\! \int\!\!\! dx_2
\delta\! \biggl[x\!-\!{M_Q x_1\!\!+\! m_q x_2 \over M_Q\!+\!m_q}\! \biggr]
     f\! \biggl[ (\!x_1\!-\!x_2\!)^2 \!\biggr]\!
\bar Q(x_1) \Gamma_H \lambda_H q(x_2)\!+\!{\rm h.c.}
\end{equation}
Here, $\lambda_H$ and $\Gamma_H$ are the Gell-Mann and Dirac matrices,
respectively, which provide the flavor and spin numbers of mesons $H$.
The function $f(x^2)$ characterizes the smearing of interaction
between a hadron and quarks.

The coupling constants $g_H$ defined by what is usually called the
{\it compositeness  condition} which means that the renormalization
constant of the meson field is equal to zero:
\begin{equation}
\label{z_h}
Z_H=1-3g^2_H/(2\pi)^2\tilde\Pi^\prime_H(m^2_H)=0
\end{equation}
where $\tilde\Pi^\prime_H$ is the derivative of the meson mass operator.

It is readily seen that in the heavy quark limit $M_Q\gg m_q$
the  Eq.(\ref{la}) becomes as
\begin{equation}
\label{lag}
{\cal L}_{\rm int} (x)\Rightarrow g_H H(x)\bar Q(x) \Gamma_H\lambda_H
\int\!\! dx_2 f\biggl[ (x-x_2)^2 \biggr] q(x_2)+{\rm h.c.}
\end{equation}
which means the  light degrees of freedom
are factorized out from the heavy ones
in according to the heavy quark symmetry. One has to remark that
the function $f(x^2)$ characterizing the long distance contributions
is related to the light quark only in this limit. It leades to
the modification of the light quark propagator in the momentum space:
\begin{equation}
{1\over m_q-\not\!p} \Longrightarrow
{f^2(p^2)\over m_q-\not\!p}
\end{equation}
In the paper \cite{HOLDOM} the monopole function $f(p^2)=1/(\Lambda^2-p^2)$
has been used for calculations of physical values. But in this case
the physical amplitudes had the threshold corresponding to quark production.
In our approach \cite{EI} an entire (nonpole) function has been used
for a single light quark propagator to ensure the quark confinement.
\begin{equation}
\label{averaging}
{1\over m_q-\not\!p}\Rightarrow
\int\!\!{d\sigma_z\over\Lambda z-\not\! p}=G(\not\! p)=
{1\over\Lambda}\biggl[a(-{p^2\over\Lambda^2})+
{\not\! p\over\Lambda} b(-{p^2\over\Lambda^2})\biggr]
\end{equation}
with the functions $a$ and $b$ being defined by
\begin{equation}
\label{fk}
a(-p^2)=\int\!\!{zd\sigma_z\over z^2-p^2}
\hspace{1.5cm}
b(-p^2)=\int\!\!{d\sigma_z\over z^2-p^2}.
\end{equation}

To conserve the local properties of Feynman diagrams like the Ward
identities, we had used the following prescription for the modification
of a line with n-light quarks within the Feyman diagram:
\begin{equation}
\label{pod}
\prod\limits_{i=0}^n \frac{1}{m_q-\not\! p_i}\Gamma_i
\Rightarrow
\int\!\! d\sigma_z \prod\limits_{i=0}^n \frac{1}{\Lambda z-\not\! p_i}\Gamma_i
\end{equation}

In QCM we had used a simple choice of the confinement
functions \cite{EI}
\begin{equation}
a(u)=a_0\exp(-u^2-a_1u) \hspace{1cm} b(u)=b_0\exp(-u^2+b_1u).
\end{equation}
The parameters $a_i$, $b_i$, and $\Lambda$ have been determined from the best
model description of hadronic properties at low energies and the following
values were found \cite{EI}:
$$a_0=b_0=2 \hspace{1cm} a_1=1 \hspace{1cm} b_1=0.4 \hspace{1cm}
{\rm and} \hspace{1cm} \Lambda=460 \; {\rm MeV},$$
which describe various physical observable quite well \cite{EI}.

Since a heavy quark (with mass $M_Q$) in a heavy meson is under the influence
of soft gluons (which sets the scale $\Lambda_{QCD} << M_Q$), it
may be regarded as nearly on its mass-shell where
the infrared regime should take place for its propagation.
The infrared behavior for one-fermion Green's function
(propagator) has been investigated in various papers (see, for instance,
\cite{KARA} and
the references therein).
The result is well-known only for abelian gauge theories:
\begin{equation}
\label{prinf}
S(p)\sim (m-\not\! p)^{-1-\nu},
\end{equation}
where $\nu=(\alpha_S/4\pi)(3-\lambda)$ with $\lambda$ being the gauge
parameter.

We have chosen \cite{INFRA} the heavy quark propagator
as in Eq.(\ref{prinf}) to calculate the weak decay constants and
the Isgur-Wise function.

 But as it is readily seen the using of such kind of propagators
in the Feynman diagrams destroys some local relations like
Ward identity which based on the obvious equality
$$
\frac{1}{m-\not\!k-\not\!p}\not\!p \frac{1}{m-\not\!k}=
\frac{1}{m-\not\!k-\not\!p}-\frac{1}{m-\not\!k}
$$

In this paper we suggest to take into account the infrared behavior
of the heavy quarks by slightly different way which nevertheless
allows one to conserve the local properties of the Feynman diagrams.
We shall use the infrapropagator for a single heavy quark line
in a diagram:
\begin{eqnarray}
\label{ip1}
&&S^h(\not\!k+\not\!p)={\Lambda^{\nu}\over (M_Q-\not\! k-\not\! p)^{1+\nu}}
=-\Lambda^{\nu}{\sin(\pi\nu) \over \pi\nu}
\int\limits_0^\infty  {du \over u^\nu} {\partial \over \partial u}
{1\over M_Q+u-\not\! k-\not\! p}
\nonumber\\
&&=\int\!\! d\sigma^h_u S^h_u(\not\!p+\not\!k)
\end{eqnarray}
Here  $M_Q$ is the constituent mass of a heavy quark and
\begin{eqnarray}
\label{ip2}
d\sigma^h_u=-{\Lambda}^\nu{\sin(\pi\nu) \over \pi\nu}
{du \over u^\nu} {\partial \over \partial u}\;\;\; ; \;\;\;
S^h_u(\not\!k+\not\!p)={1\over M_Q+u-\not\! k-\not\! p}
\end{eqnarray}

But we will use the prescription for a line with n heavy quarks analogous
to those used in Eq.(\ref{pod}) for averaging of light quarks:
\begin{equation}
\label{ip3}
\prod\limits_{i=0}^n \frac{1}{M_{Q_i}-\not\! p_i}\Gamma_i
\Rightarrow
\int\!\! d\sigma^h_u \prod\limits_{i=0}^n
\frac{1}{M_{Q_i}+u-\not\! p_i}\Gamma_i
\end{equation}

Let $K^{(0)}(M_{Q_1},...,M_{Q_n})$ be the structural integral corresponding to
the Feynman diagram with n heavy quark local propagators. It is readily
seen that the taking into account the infrared behavior according to
Eqs.(\ref{ip1},\ref{ip2},\ref{ip3}) gives
\begin{eqnarray}
\label{sigm}
&&K^{(\nu)}(M_{Q_1},\!...,M_{Q_n}\!)\!=
\!\int\!d\sigma^h_u K^{(0)}(M_{Q_1}\!+\!u,\!...,M_{Q_n}\!+\!u)
\nonumber\\
&&=\!-{\sin(\pi\nu) \over \pi\nu}
\!\!\int\limits^\infty_0\!\!\!{du \over u^\nu}{\partial \over \partial u}
K^{(0)}(\!M_{Q_1}\!+\!u,\!...,M_{Q_n}\!+\!u\!)
\!\!=\!\!{\sin(\!\pi\nu\!) \over \pi\nu}\!\biggl\{\!K^{(0)}(\!M_{Q_1},\!...,M_{Q_n}\!)
\nonumber\\
&&+\nu
\int\limits^1_0\!\! {du \over u^\nu}{K^{(0)}(M_{Q_1},\!...,M_{Q_n})-K^{(0)}(M_{Q_1}+u,\!...,M_{Q_n}+u)\over u}
\nonumber\\
&&-\nu\int\limits^\infty_1\!\!
{du \over u^{\nu+1}}K^{(0)}(M_{Q_1}+u,\!...,M_{Q_n}+u\!) \biggr\}
\!\approx\!
{\sin(\pi\nu) \over \pi\nu}\biggl\{K^{(0)}(M_{Q_1},...,M_{Q_n})
\nonumber\\
&&+\nu
\int\limits^1_0\! {du \over u^\nu}{K^{(0)}(M_{Q_1},...,M_{Q_n})-K^{(0)}(M_{Q_1}+u,...,M_{Q_n}+u)\over u}
\biggr\}
\end{eqnarray}
This  means that we can calculate the integral corresponding to
Feynman diagram with the local heavy quark propagators  and then
apply the averaging according to Eq.(\ref{sigm}) to take into
account the infrared behavior.

\section{Heavy-to-light form factors}

First, we introduce the necessary notation. We will write $P$ for
light pseudoscalar meson and $V$ for light vector meson with the masses
$m_P$ and $m_V$ respectively. We will write $H$ for heavy pseudoscalar meson
with a  mass $m_H$.

In what follows we assume that all masses and momenta in the structural
integrals are given in units of $\Lambda=460 \; {\rm MeV}$.

The invariant amplitudes for $H\to l\nu$ and $H\to P(V)l\nu$ transitions are
defined as
\begin{eqnarray}
&&A(H\to e \nu)=
{ G \over \sqrt{2} }
V_{qq^\prime}
(\bar e O_{\mu}\nu) M_H^\mu(p)\\
&&A(H\to Pe\nu)={G\over \sqrt{2}}V_{qq^\prime}(\bar eO_{\mu}\nu)
M^{\mu}_{HP}(p,p^\prime)\\
&&A(H\to Ve\nu)={G\over \sqrt{2}}V_{qq^\prime}(\bar eO_{\mu}\nu)\epsilon_\nu
M^{\mu\nu}_{HV}(p,p^\prime),
\end{eqnarray}
where
\begin{eqnarray}
\label{i3}
&&\hspace*{-.6cm}M_H^\mu(p)=
{3\over 4\pi^2}g_H\!\!\int\!\! d\sigma^h_u\!\!\int\!\! d\sigma^l_z
\!\!\int\!\!{d^4k\over 4\pi^2i}
{\rm tr}\biggl\{i\gamma^5 S^h_u(\not\! k+\not\! p)
O^\mu S^l_z(\not\! k)
\biggr\}=i  f_H p^\mu\\
\nonumber\\
\label{i4}
&&\hspace*{-.6cm}M^{\mu}_{HP}(p,p^\prime)\!=\!\!
{3\over 4\pi^2}g_Hg_P\!\!\!\int\!\!\! d\sigma^h_u\!\!\!\int\!\!\! d\sigma^l_z
\!\!\!\int\!\!\!{d^4k\over 4\pi^2i}
{\rm tr}\biggl\{\!i\gamma^5 S^h_u(\!\not\! k+\!\not\! p)
O^\mu S^l_z(\!\not\! k+\!\not\! p^{\,\prime})i\gamma^5 S^l_z(\!\not\! k)\!\biggr\}
\nonumber\\
&&\hspace*{-.6cm}=f^+(q^2)(p^\mu+p^{\prime\mu}) + f^-(q^2)(p^\mu-p^{\prime\mu})\\
\nonumber\\
\label{i5}
&&\hspace*{-.6cm}M^{\mu\nu}_{HV}(p,p^\prime)\!=\!\!
{3\over 4\pi^2}g_Hg_V\!\!\!\int\!\!\! d\sigma^h_u\!\!\!\int\!\!\! d\sigma^l_z
\!\!\!\int\!\!\!{d^4k\over 4\pi^2i}
{\rm tr}\biggl\{\!i\gamma^5 S^h_u(\!\not\! k+\!\not\! p)
O^\mu S^l_z(\!\not\! k+\!\not\! p^{\,\prime})\gamma^\nu S^l_z(\!\not\! k)\!\biggr\}
\nonumber\\
&&\hspace*{-.6cm}\!=\!-i(m_H\!+\!m_V) A_1(q^2)g^{\mu\nu}
\!+\!i{ A_2(q^2) \over m_H\!+\!m_V}(p^\mu\!+\!p^{\prime\mu})p^\nu
\!+\!i{ A_0(q^2) \over m_H\!+\!m_V}(p^\mu\!-\!p^{\prime\mu})p^\nu\nonumber\\
&&\hspace*{-.6cm}+{2 V(q^2) \over m_H+m_V} e^\mu_{\nu\alpha\beta}
p^\alpha p^{\prime\beta}
\end{eqnarray}
Here
\begin{eqnarray}
q^2=(p-p^\prime)^2,\;\;\;\;\;\;p^2=m^2_H,\;\;\;\;\;\;p^{\prime\,2}=m^2_{P(V)}
\end{eqnarray}
The integrals like (\ref{i3}), (\ref{i4}), (\ref{i5}) are calculated by using
the standard Feyman $\alpha$ -- para\-metri\-zation and the prescriptions
(\ref{fk}) and (\ref{sigm}) to integrate over $d\sigma^l_z$ and
$d\sigma^h_u$, respectively.
The explicit expressions for the form factors are given in Appendix.

The hadron-quark coupling constants for light pseudoscalar and vector
mesons and heavy pseudoscalar mesons determined from the compositeness
condition (\ref{z_h}) are written down
\begin{eqnarray}
g_P={2\pi \over \sqrt 3}\sqrt{2 \over R_P(m_P^2)},\quad\quad
g_V=2\pi \sqrt{1 \over R_V(m_V^2)},\quad\quad
g_H={2\pi \over \sqrt 3}\sqrt{1 \over  R_H(m^2_H)}
\end{eqnarray}
The functions $R_i(x)$ are shown in Appendix.

\section{Numerical results and discussions}

We have two adjustable parameters in our approach: the difference
between the heavy meson and heavy quark masses  $E=m_D-M_c=m_B-M_b$
and the parameter $\nu$ characterizing the infrared behavior
of heavy quark. We will adjust them by fitting the available
experimental data for branching ratios of D-meson decays.
The results of fit are given in Table 1. One has to remark that
the allowed regions for the adjustable parameters are quite narrow:
the parameter E varies from 0.315 up to 0.415 GeV whereas
the infrared parameter $\nu$ varies from 0.6 up 0.7. In what follows
we will give our results for E=0.365 GeV and $\nu$=0.65. This set
of the parameters seems to be best for describing the available
experimental data. Our results for $f_D$ and $f_B$ in this case are
\begin{equation}
 f_D=193\;{\rm MeV},\;\;\;\;\;\;f_B=136\;{\rm MeV}
\end{equation}

The magnitudes of form factors of the $D\to K(K^*)e\nu$  at $q^2=0$ are given
in Table 2.  One can see that our results are
in good agreement with the experimental data and other approaches.
The dependence on the momentum transfer squared  is shown in Fig.1 and
Fig.2.

The results for the form factors and decay widths of B-meson
are given in Tables 3 and 4 by using the values of adjustable parameters
obtained above. It is seen that the values of the form factors at
$q^2=0$ are substantially larger those obtained in QCD Sum Rules \cite{N}--
\cite{BBD} and quark models \cite{FGM}-\cite{BSW} but close to the results
of papers \cite{GSS}, \cite{DP}, \cite{Bec}, \cite{Cas} and \cite{Ch} for
$f^+(0)$. In the latter it was found that using the heavy quark symmetry
and assuming pole dominance for the form factors, $f_+^{B\pi}(0)$ is
estimated to be $\approx$ 0.39.
If the requirement of heavy quark symmetry is weaken so that it applies only
to soft pion emissions from the heavy meson, one finds
$f_+^{B\pi}(0)\approx 0.53$.

Recently, the CLEO collaboration \cite{cleo} claims the following results
for semileptonic $B$ decays branching ratios and $|V_{ub}|$:
\begin{eqnarray}\label{cleo}
&&{\rm Br}(B^0 \to \pi^- l^+ \nu) =
(1.8\pm 0.4\pm 0.3\pm 0.2)\times 10^{-4}\nonumber\\
&&{\rm Br}(B^0 \to \rho^- l^+ \nu) =
(2.5\pm 0.4^{+0.5}_{-0.7}\pm 0.5)\times 10^{-4}\\
&&|V_{ub}|=
(3.3\pm 0.2^{+0.3}_{-0.4}\pm 0.7)\times 10^{-3}, \nonumber
\end{eqnarray}
One has to note that the branching fractions for
$B \to \pi e \nu$ and $B \to \rho e \nu$ have been obtained by using a variety
of models. What is given in Eq.~(\ref{cleo}) is just the average value
of the branching fractions obtained for each model. The errors are
statistical, systematic and the estimated model dependence.
The values for $|V_{\rm ub}|$ were extracted from the branching fractions
using $\tau_{B_0}=1.56\pm 0.05$ ps. To obtain $|V_{\rm ub}|_{\rm aver}$,
the $\pi$ and $\rho$ modes were combined by fixing their ratio to
the prediction for each model.

This means that it is not easy to use these results for checking predictions
of other models. Nevertheless, we extract the values
$\Gamma(B\to \pi(\rho)e\nu)/|V_{ub}|^2\times 10^{12} s^{-1}$ from
Eq.~(\ref{cleo}) and put them without errors in Table 4.
As one expects our predictions are almost twice larger those
values obtained in a such manner.  At the same time the ratio $\pi$/$\rho$
is in a good agreement with the experimental result.

The $q^2$ dependence of form factors is shown in Fig. 3.
        
One can see from Fig.1-3 that the behavior of the form factors
$f^+(q^2)$ and $V(q^2)$ in the physical regions of the momentum transfer squared
($q^2<1.5\;{\rm Gev^2}$ for the D-meson decay and $q^2<20.0\;{\rm Gev^2}$ for
the B-meson decay) agrees quite well with $q^2$ dependence of the
monopole form factor. At the same time the behavior of the form factors
$A_1(q^2)$ and $A_2(q^2)$ is rather different from the monopole function and
similar to the prediction of QCD sum rules \cite{BBD}.

\section{Conclusion}

We have slightly modified our previous work \cite{INFRA} to  take into
consideration the infrared behavior of the heavy quark in consistent way to
provide  the conservation of local properties of Feynman diagrams like
gauge invariance and the Ward-Takachashi  identity.
The infrared behavior of the heavy quark is modelled in
terms of a single parameter $\nu$ which modifies the simple free Feynman
propagator.

We have investigated the dependence of the heavy-to-light form factors
calculated within this approach on two parameters:
the infrared parameter $\nu$ and the mass difference
of the heavy meson and heavy quark $E=m_H-M_Q$.

We  found that the values of the $D\to K(K^*)$ or $D\to \pi(\rho)$
form factors at $q^2=0$ are in excellent agreement with the available
experimental data and other approaches whereas these values for
$B\to\pi(\rho)$ transitions are found to be larger than those of
several other models.

We have used the obtained form factors to calculate the widths of
the semileptonic decays of B and D mesons.

\begin{acknowledge}
We  would like to thank B.Stech for useful discussions.
This work was supported in part by  the United States Department of Energy
under Grant No. DE-FG05-84-ER40413, by the INTAS Grant 94-739 and 
by the Russian Fund of Fundamental Research (RFFR) under 
contract 96-02-17435-a.
\end{acknowledge}

\hspace{1cm}

\small
\section*{Appendix}

\noindent
{\it Two-point function:}
\begin{eqnarray}
f_{P}&=&
g_P\,\,{3\Lambda \over (2\pi)^2}
\biggl\{ A_0+
{m^2_P \over 4} \int\limits^1_0\!\! d\alpha\;
{\bf a}\biggl(-\alpha{m^2_P \over 4}\biggr)\sqrt{1-\alpha} \biggr\},
\nonumber\\
R_P(x)&=&B_0+{x\over 4}\int\limits_0^1\!\! du b(-{ux\over 4})
{(1-u/2)\over\sqrt{1-u}},
\nonumber\\
R_V(x)&=&B_0+{x\over 4}\int\limits_0^1\!\! du b(-{ux\over 4})
{(1-u/2+u^2/4) \over \sqrt{1-u}},
\nonumber\\
R_H(x)&=&\int\!\! d\sigma^h_u \!\! \int\limits_0^1\!\! d\alpha \alpha
\biggl[M {\bf a}\biggl(\delta(\alpha,x)\biggr)
+\biggl({3\alpha-2\alpha^2 \over 2(1-\alpha)^2}M^2
-{1\over2}\alpha x\biggr)
{\bf b}\biggl(\delta(\alpha,x)\biggr)\biggr]
\nonumber
\end{eqnarray}

\noindent
{\it Heavy-to-light form factors:}
\begin{eqnarray}
f_H\!\!\!&=&\!\!\! g_H{3\Lambda \over (2\pi)^2}\!\!
\int\!\!\! d\sigma^h_u\!\!\! \int\limits^1_0\!\!\! d\alpha \alpha\!
\biggl(\!{M^2\over (\!1\!-\!\alpha)^2}
\!-\!m_H^2\!\biggr)\!\!\biggl[\!(1\!-\!{\alpha \over 2})
{\bf a}\!\biggl(\!\!\delta(\alpha,m_H^2)\!\!\biggr)\!\!+\!\!M{\alpha\over 2}
{\bf b}\!\biggl(\!\!\delta(\alpha,m_H^2)\!\!\biggr)
\!\!\biggr]
\nonumber\\
&&\nonumber\\
f^+(q^2)\!\!\!&=&\!\!\!
g_Hg_P\! {3 \over 2(2\pi)^2}\!\!\! \int\!\!\! d\sigma^h_u
\biggl\{
\!\! B_0\!\!+\!\! {m^2_P \over 4}\!\!\! \int\limits^1_0\!\!\! d\alpha
{\bf b}\!\biggl(\!\!-\alpha{m^2_P \over 4}\!\!\biggr)\!\sqrt{1\!-\!\alpha}\!
\biggr.
+\!\!\int\limits_0^1 \!\!\!d\alpha_1\!\!\! \int\limits_0^1\!\!\! d\alpha_2
\!\biggl[\! M{\bf a}\!\biggl(\!\!\Delta(\!\alpha_1,\alpha_2\!)\!\!\biggr)
\nonumber\\
&+&
\biggl.
\!\!\!\biggl((1-\alpha_1)(1-\alpha_2)m^2_H-M^2+
\alpha_2(1-\alpha_1)q^2\biggr)
{\bf b}\biggl(\Delta(\alpha_1,\alpha_2)
\biggr) \biggr]\biggr\}
\nonumber\\
&&\nonumber\\
f^-(q^2)\!\!\!&=&\!\!\!-f^+(q^2)
+g_Hg_P {3 \over (2\pi)^2}\int\!\!  d\sigma^h_u
\biggr\{\!\!
\int\limits_0^1 \!\!d\alpha_1\!\! \int\limits_0^1\! \!d\alpha_2\,
\alpha_2(1-\alpha_1)m^2_P
{\bf b}\biggl(\!\Delta(\alpha_1,\alpha_2)\!\biggr)
\biggr.
\nonumber\\
&+&
\biggl.
\int\limits^1_0\! d\alpha\, \alpha
(1-{\alpha \over 2})
\biggl({M^2\over (1-\alpha)^2}
-q^2\biggr)
{\bf b}\biggl(\delta(\alpha,q^2)\biggr)
\biggr\}
\nonumber\\
&&\nonumber\\
A_1(q^2)\!\!\!&=&\!\!\!g_Hg_V
{3 \over (2\pi)^2(\!m_H\!+\!m_V\!)}\int\!\!  d\sigma^h_u
\biggr\{
\int\limits^1_0\! d\alpha \alpha
\biggl({M^2\over (1-\alpha)^2}
-q^2\biggr)
{\bf b}\biggl(\delta(\alpha,q^2)\biggr)
\nonumber\\
&+&\!\!\int\limits_0^1\! d\alpha_1 \int\limits_0^1\! d\alpha_2
{m^2_H+m^2_V-q^2 \over 2}
{\bf a}\biggl(\Delta(\alpha_1,\alpha_2)\biggr)
\nonumber\\
&+&\!\!\!\int\limits_0^1 \!\!\!d\alpha_1\!\!\! \int\limits_0^1 \!\!\!d\alpha_2
\!\biggl[\!\alpha_1\!\biggl(\!\!1\!-\!{1\over 2}\alpha_1\!\!\biggr)\!\!\biggl(\!{M^2
\over (\!1\!-\!\alpha_1\!)^2}
\!-\!m_H^2\!\biggr)\!-\!\alpha_1\!\biggl(\!\!1\!-\!2\alpha_2\!+
\!\alpha_1\alpha_2\!\!\biggr)\!
{m^2_H\!+\!m^2_V\!-\!q^2 \over 2}
\nonumber\\
&-&\!\!\!\alpha_2\biggl(1\!-\! \alpha_1\!+
\!\alpha_1\alpha_2\!-\!{1\over 2}\alpha_1^2\alpha_2\biggr)\!\biggr]
\biggl[{\bf a}\biggl(\!\Delta(\!\alpha_1,\alpha_2\!)\!\biggr)
-M{\bf b}\biggl(\!\Delta(\!\alpha_1,\alpha_2\!)\!\biggr)
\!\biggr]\!\biggr\}
\nonumber\\
&&\nonumber\\
A_2(q^2)\!\!\!&=&\!\!\!g_Hg_V\!
{3(\!m_H\!+\!m_V\!) \over 2(2\pi)^2}\!\!\! \int\!\!\! d\sigma^h_u\!\!\!
\int\limits_0^1\!\!\!d\alpha_1\!\!\! \int\limits_0^1\!\!\! d\alpha_2 \!\biggr[
\!\biggr(\!\!1\!\!-\!\!3\alpha_1\!\!+\!\!2\alpha_1\alpha_2\!\!
+\!\!2\alpha_1^2\!\!-\!\!2\alpha_1^2\alpha_2\!\!\biggr)
{\bf a}\!\biggl(\!\!\Delta(\!\alpha_1,\!\alpha_2\!)\!\!\biggr)
\nonumber\\
\!\!\!&+&\!\!\!\alpha_1
\biggl(1-2\alpha_1-2\alpha_2+2\alpha_1\alpha_2)\biggr)M
{\bf b}\biggl(\Delta(\alpha_1,\alpha_2)\biggr)
\biggr]
\nonumber\\
&&\nonumber\\
A_0(q^2)\!\!\!&=&\!\!\!-\! A_2(q^2)\!-\!g_Hg_V\!
{3(\!m_H\!+\!m_V\!) \over (2\pi)^2}\!\!\int\!\!  d\sigma^h_u\!\!
\int\limits_0^1 \!\!d\alpha_1\!\! \int\limits_0^1\!\! d\alpha_2 \biggr[
2\alpha_1(1-\alpha_1)\,
{\bf a}\!\biggl(\!\Delta(\!\alpha_1,\alpha_2\!)\!\biggr)
\nonumber\\
\!\!\!&+&\!\!\!2\alpha_1^2M
{\bf b}\biggl(\Delta(\alpha_1,\alpha_2)\biggr)
\biggr]
\nonumber\\
&&\nonumber\\
V(q^2)\!\!\!&=&\!\!\!g_Hg_V\!{3(\!m_H\!+\!m_V\!) \over 2(2\pi)^2}\!\!
\int\!\!\!  d\sigma^h_u \!\!\!
\int\limits_0^1 \!\!\!d\alpha_1\!\!\! \int\limits_0^1\!\!\! d\alpha_2
\biggr[
\!\alpha_1M {\bf b}\!\biggl(\!\!\Delta(\!\alpha_1,\!\alpha_2\!)\!\!\biggr)
\!+\!(1\!-\!\alpha_1) {\bf a}\!\biggl(\!\!\Delta(\!\alpha_1,\!\alpha_2\!)\!\!\biggr)
\!\biggr]
\nonumber
\end{eqnarray}

\noindent
{\it Here}
\begin{eqnarray}
&&\hspace*{-1.2cm} M=M_Q+u \nonumber\\
&&\hspace*{-1.2cm}A_n=\int\limits_0^\infty\!\! du a(u)u^n,\quad\quad
B_n=\int\limits_0^\infty\!\! du b(u)u^n,\quad\quad
\delta(\alpha,x)=
{\alpha\over(1-\alpha)}M^2-\alpha x\nonumber\\
&&\hspace*{-1.2cm}\Delta(\alpha_1,\alpha_2)=
{\alpha_1\over (1-\alpha_1)}M^2
-\alpha_1(1-\alpha_2)m_H^2
-(1-\alpha_1)(1-\alpha_2)\alpha_2 m_{P(V)}^2
-\alpha_1\alpha_2 q^2\nonumber
\end{eqnarray}

\newpage

\begin{table}
\caption[t1]{Branching ratios for semileptonic and
nonleptonic B and D meson decays for various values of the model
parameters $E$ and $\nu$.}
\begin{tabular}{lcccc}
\hline
${\rm Decay}$
&${\rm Experimental}$
&$E=0.415$
&$E=0.365$
&$E=0.315$\\
&$\!{\rm average\;\cite{EXP}}\!$
&$\nu=0.60$
&$\nu=0.65$
&$\nu=0.70$\\
\hline
$Br(D^0\!\! \to\!\! K^- e^+ \nu_e)$      &$3.8 \pm 0.22$          &4.26 &3.77 &3.32\\
$Br(D^0\!\!\to\!\!\pi^-e^+\nu_e)$        &$0.39^{+0.23}_{-0.12}$  &0.48 &0.42 &0.35\\
\hline
$Br(D^0\!\! \to\!\! K^{*-} e^+ \nu_e)$   &$2.0 \pm 0.4$           &1.39 &1.31 &1.23\\
\hline
$Br(D^+\!\!\to\!\!\bar{K}^0e^+\nu_e)$    &$6.6 \pm 0.9$           &11.0 &9.82 &8.57\\
$Br(D^+\!\!\to\!\!\pi^0l^+\nu_l)$        &$0.57 \pm 0.22$         &0.63 &0.55 &0.47\\
\hline
$Br(D^+\!\!\to\!\!\bar{K}^{*0}e^+\nu_e)$ &$4.8 \pm 0.5$           &3.54 &3.34 &3.14\\
$Br(D^+\!\!\to\!\!\rho e^+\nu_e)$        &$< 0.37$                &0.35 &0.32 &0.29\\
\hline
\end{tabular}
\end{table}

\begin{table}
\caption[t2]{The predictions for the form factors at $q^2=0$
in the decay $D\to K(K^*) e\nu$ along with those of other approaches
and the experiment.}
\begin{tabular}{lcccc}
\hline
$\!\!D\!\!\to\!\! K(\!K^*\!)e\nu_e\!\!$         &$f^+(0)$               &$A_1(0)$              &$A_2(0)$                     &$V(0)$\\
\hline
$\!\!{\rm Expt.average}\!\!$                    &$0.75\!\pm\! 0.04$     &$0.56\!\pm\! 0.04$    &$0.4\! \pm\! 0.08$        &$1.1 \!\pm\! 0.2$\\
\hline
${\rm Our\;results}$                            &0.78                   &0.52                  &0.4                                     &1.07\\
\hline
${\rm BKS \cite{BKS}}$                           &$0.90^{\pm 8}_{\pm 21}$  &$0.83^{\pm 14}_{\pm 28}$ &$0.59^{\pm 14}_{\pm 24}$
 &$1.43^{\pm 45}_{\pm 49}$\\
${\rm LMMS \cite{LMMS}}$                         &$0.63\! \pm\! 0.08$  &$0.53\! \pm \!0.03$ &$0.19\! \pm\! 0.21$   &$0.86\!\pm\! 0.1$\\
${\rm ELC \cite{Aba}}$                           &$0.60^{\pm 15}_{\pm 7}$  &$0.64\!\pm\!0.16$ &$0.4^{\pm 28}_{\pm 4}$  &$\!0.86\!\pm\!0.24\!$\\
${\rm APE \cite{APE}}$                           &$0.78\!\pm\!0.08$  &$0.67\!\pm\!0.11$ &$\!0.49\!\pm\!0.34\!$ &$\!1.08\!\pm\!0.22\!$\\
${\rm UKQCD \cite{UKQCD}}$                       &$0.67^{+7}_{-8}$  &$0.70^{+7}_{-10}$ &$0.66^{+10}_{-15}$ &$1.01^{+30}_{-13}$\\
${\rm GSS \cite{GSS}}$                           &$\!0.71^{+12\;+10}_{-12\;-7}\!$  &$\!0.61^{+6\;+9}_{-6\;-7}\!$ &$\!0.83^{+20\;+12}_{-20\;-8}\!$  &$\!1.34^{+24\;+19}_{-24\;-14}\!$\\
${\rm DP  \cite{DP}}$                            &$0.75\!\pm\!0.05$     & & &\\
${\rm BBD \cite{BBD}}$                           &$0.60^{+15}_{-10}$  &$0.5\!\pm\!0.15$  &$0.6\!\pm\!0.15$   &$1.1\!\pm\!0.25$\\
${\rm ISGW \cite{ISGW}}$                         &$0.76\!-\!0.82$      &0.8             &0.8       &1.1\\
${\rm FGM \cite{FGM}}$                           &$0.73\! \pm\! 0.07$  &$0.63\! \pm\! 0.06$ &$0.43\! \pm\! 0.04$    &$0.62\! \pm\! 0.06$\\
${\rm BD \cite{BD}}$                             &0.8  &0.77 &1.48 &1.32\\
${\rm AW \cite{AW}}$                             &0.7              &0.8             &0.6              &1.5\\
${\rm BSW \cite{BSW}}$                           &0.76             &0.88            &1.15             &1.3\\
${\rm J \cite{Wol}}$                             &0.78 &0.66 &0.43 &1.04\\
${\rm C\cite{Cas}}$                              &0.67  &0.48 &0.27 &0.95\\
\hline
\end{tabular}
\end{table}

\begin{table}
\caption[t3]{The predictions for the  form factors at $q^2=0$
in the decay $B\to \pi(\rho) e\nu$ along with those of other approaches.}
\begin{tabular}{lcccc}
\hline
$\!\!B\!\!\to\!\! \pi(\rho)e\nu_e\!\!$          &$f^+(0)$  &$A_1(0)$  &$A_2(0)$  &$V(0)$\\
\hline
${\rm Our\;results}$                             &0.53 &0.50 &0.51 &0.70\\
\hline
${\rm ELC \cite{Aba}}$                           &$0.30^{\pm 14}_{\pm 5}$  &$0.22\!\pm\!0.05$ &$0.49^{\pm 21}_{\pm 5}$ &$\!0.37\!\pm\!0.11\!$\\
${\rm APE \cite{APE}}$                           &$0.35\!\pm\!0.08$  &$0.24\!\pm\!0.12$ &$\!0.27\!\pm\!0.80\!$  &$\!0.53\!\pm\!0.31\!$\\
${\rm UKQCD \cite{UKQCD}}$                       &$0.23\!\pm\!0.02$  &$0.27^{+7\;+3}_{-4\;-3}$ &$0.28^{+9\;+4}_{-6\;-5}$ &\\
${\rm GSS \cite{GSS}}$                           &$\!0.50^{+14\;+7}_{-14\;-5}\!$ &$\!0.16^{+4\;+22}_{-4\;-16}\!$ &$\!0.72^{+35\;+10}_{-35\;-7}\!$  &$\!0.61^{+23\;+9}_{-23\;-6}\!$\\
${\rm DP \cite{DP}}$                             &$0.4 \!\pm\! 0.1$   & & &\\
${\rm N \cite{N}}$                               &$0.23\!\pm\! 0.02$  &$0.38\!\pm\! 0.04$ &$\!0.45\!\pm\! 0.05\!$ &$\!0.45\!\pm\! 0.05\!$\\
${\rm CS \cite{Col}}$                            &0.24  & & &\\
${\rm B \cite{BBD}}$                             &$0.26\!\pm\! 0.02$  &$0.5\!\pm\! 0.1$   &$0.4\!\pm\! 0.2$  &$0.6\!\pm\! 0.2$\\
${\rm ISGW \cite{ISGW}}$                         &0.09 &0.05 &0.02 &0.27\\
${\rm FGM \cite{FGM}}$                           &$0.21\!\pm\! 0.02$  &$0.26\!\pm\! 0.03$ &$\!0.30\!\pm\! 0.03\!$ &$\!0.29\!\pm\! 0.03\!$\\
${\rm BSW \cite{BSW}}$                           &0.33 &0.28 &0.28  &0.33\\
${\rm J \cite{Wol}}$                             &0.27 &0.26 &0.24 &0.35\\
${\rm M \cite{Mel}}$                             &$\!0.29\!\sim\!0.2\!$  &$\!0.26\!\sim\!0.17\!$ &$\!0.24\!\sim\!0.16\!$ &$\!0.34\!\sim\!0.22\!$\\
${\rm B\cite{Bec}}$                              &$0.38\!\pm\!0.3$  & & &\\
${\rm C\cite{Cas}}$                              &0.89  &0.21 &0.2 &1.04\\
${\rm Ch.\cite{Ch}}$                             &0.53 & & &\\
\hline
\end{tabular}
\end{table}

\begin{table}
\caption[t4]{Semileptonic decay rates $\Gamma(B\to \pi(\rho)e\nu)$
in units $|V_{ub}|^2\times 10^{12} s^{-1}$.
The comparison of our results with those of other approaches.}
\begin{tabular}{lccc}
\hline
${\rm Decay}$                                   &$B\!\! \to\!\!\pi e \nu_e$  &$B\!\!\to\!\!\rho e \nu_e$
&$\Gamma(B\!\!\to\!\!\rho e \nu_e)/\Gamma(B\!\!\to\!\!\pi e \nu_e)\!\!$\\
\hline
${\rm CLEO  \cite{cleo}}$     &(10.59) &(14.72) &$1.4^{+0.6}_{-0.4}\!\pm\!0.3\!\pm\!0.4$\\
\hline
${\rm Our\;results}$                            &19.0               &25.0                &1.32\\
\hline
${\rm ELC \cite{Aba}}$                          &$9\!\pm\!6$  &$14\!\pm\!12$            &(1.55)\\
${\rm APE \cite{APE}}$                          &$8\!\pm\!4$  &                         &      \\
${\rm DP \cite{DP}}$                            &$14.5\!\pm\!0.59$  &                   &\\
${\rm N \cite{N}}$                              &$3.6\!\pm\! 0.6$   &$5.1\!\pm\! 1.0$   &(1.41)\\
${\rm B \cite{BBD}}$                            &$5.1\!\pm\! 1.1$   &$12\!\pm\! 4$      &(2.3)\\
${\rm ISGW \cite{ISGW}}$                        &2.1                &8.3                &3.9\\
${\rm FGM \cite{FGM}}$                          &$3.1\!\pm\! 0.6$   &$5.7\!\pm\! 1.2$   &(1.83)\\
${\rm BSW \cite{BSW}}$                          &7.4                &26                 &3.5\\
${\rm J \cite{Wol}}$                            &10.0  &19.1 &1.91 \\
${\rm M \cite{Mel}}$                            &$7\!\pm\! 2$       &$10.6\!\pm\! 3.2$   &$1.45\!\pm\!0.1$\\
${\rm C \cite{Cas}}$                            &54.0  &34.0 &0.63 \\
\hline
\end{tabular}
\end{table}

\begin{figure}[htb]
\mbox{\hspace*{0.7cm}\epsfysize=15.cm\epsfxsize=12.cm\epsffile{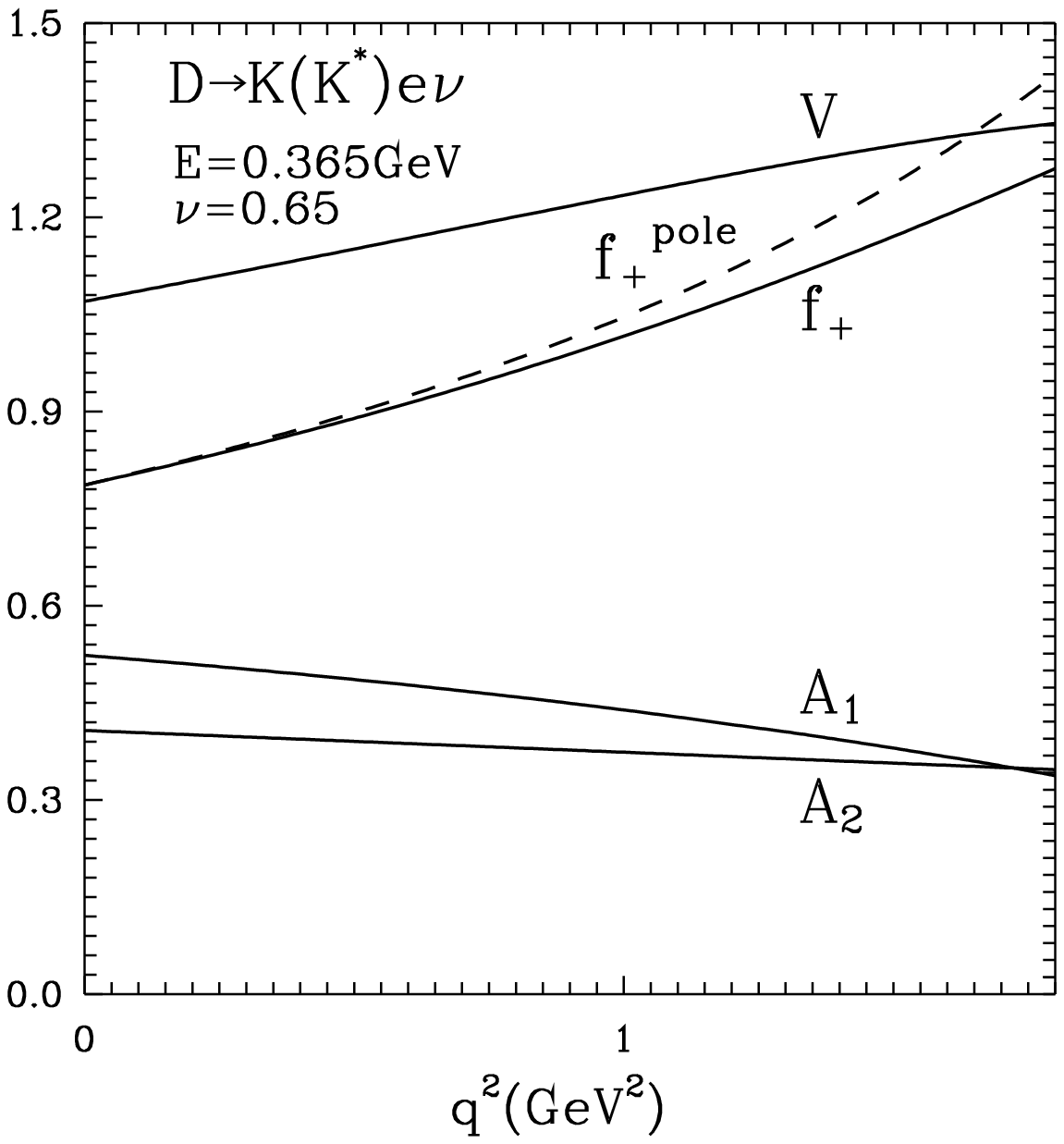}}
\vspace{-7.4cm}
\caption[f1]{$q^2$-dependence  of the form factors relevant for the decays
$D\to K(K^*) e\nu$.}
\end{figure}

\begin{figure}[p]
\mbox{\hspace*{0.7cm}\epsfysize=15.cm\epsfxsize=12.cm\epsffile{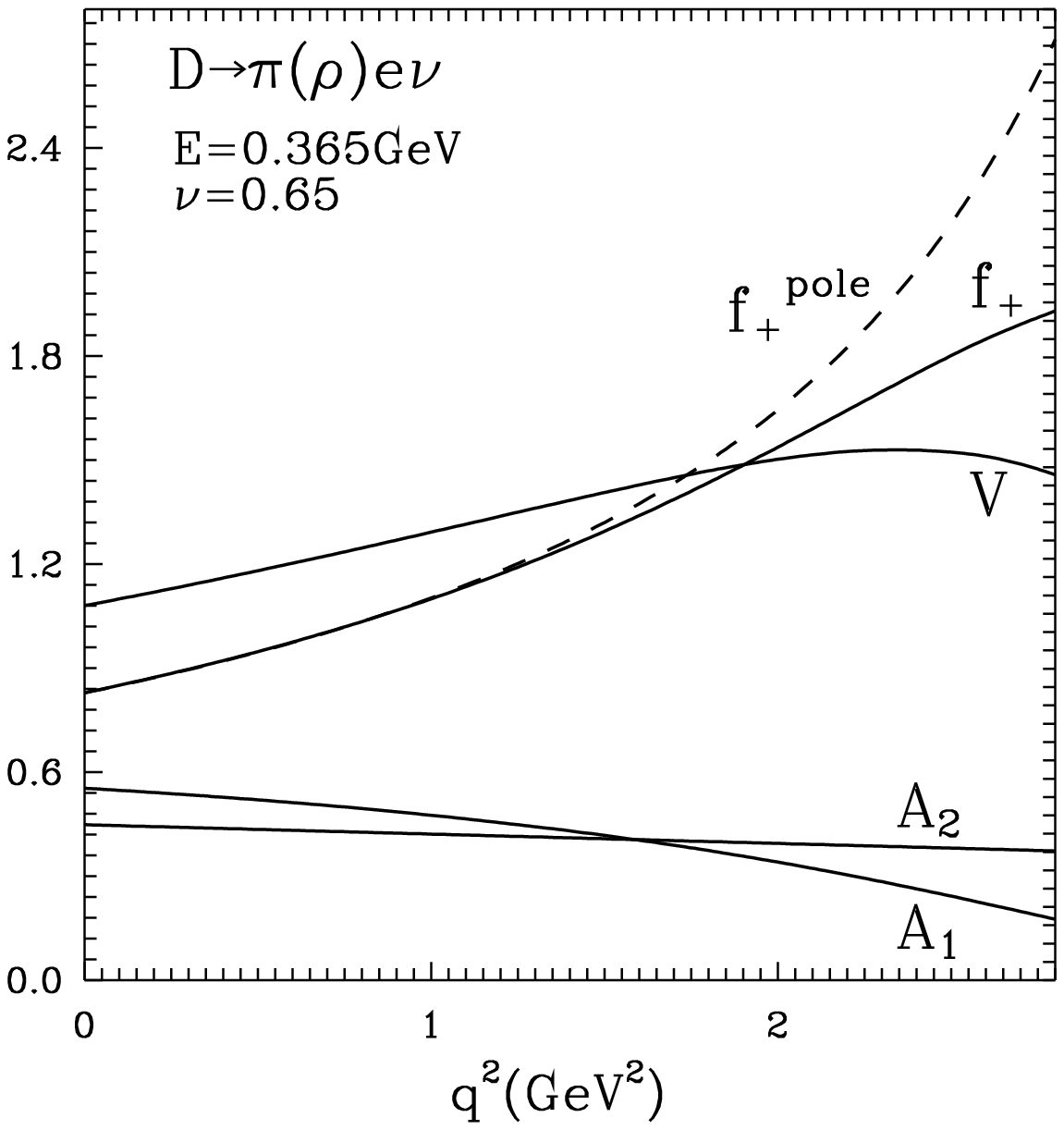}}
\vspace{-7.4cm}
\caption[f2]{$q^2$-dependence  of the form factors relevant for the decays
$D\to \pi(\rho) e\nu$.}
\end{figure}

\begin{figure}[p]
\mbox{\hspace*{0.7cm}\epsfysize=15.cm\epsfxsize=12.cm\epsffile{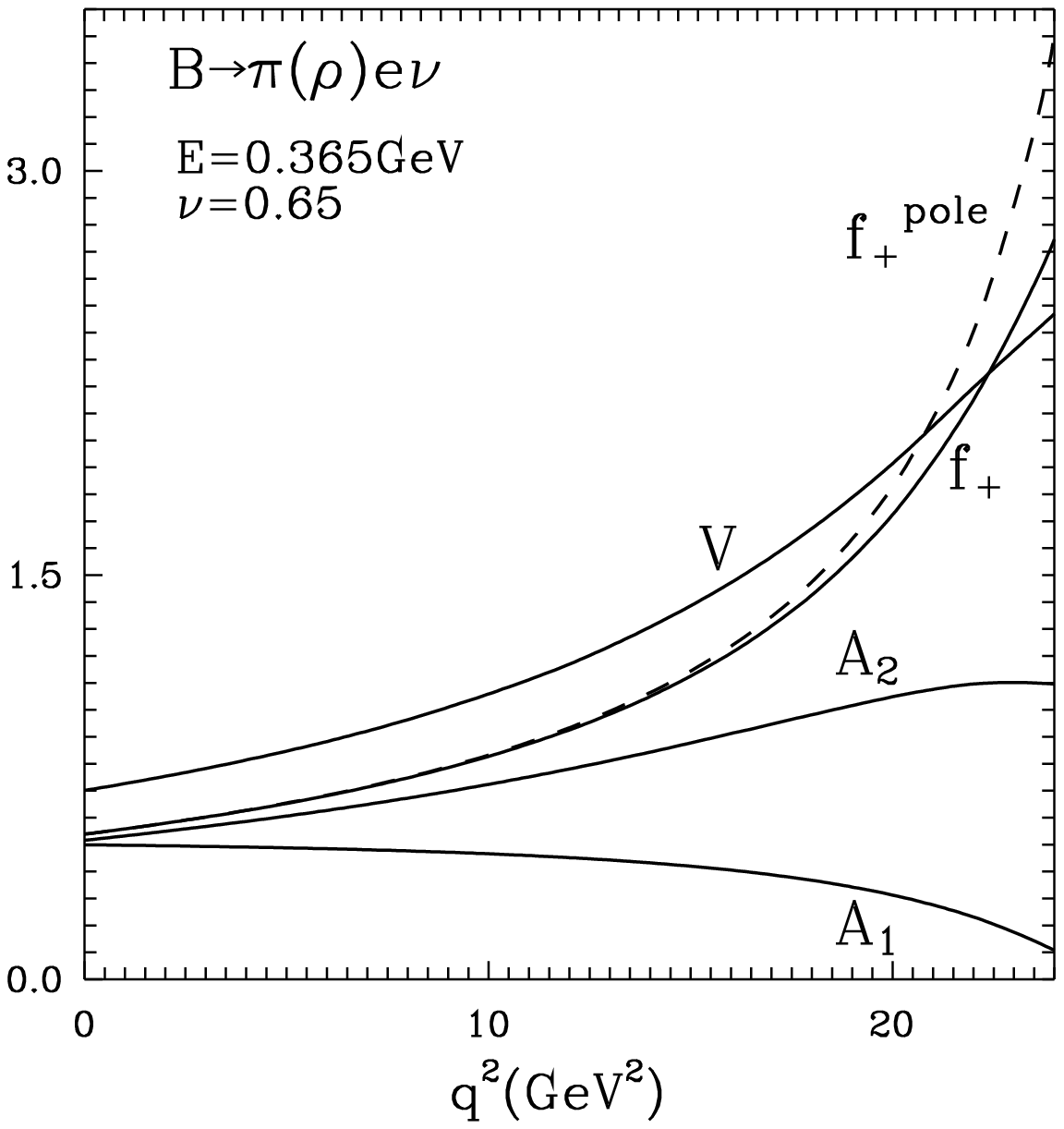}}
\vspace{-7.4cm}
\caption[f3]{$q^2$-dependence  of the form factors relevant for the decays
$B\to \pi(\rho) e\nu$.}
\end{figure}

\SaveFinalPage
\end{document}